\UseRawInputEncoding

\documentclass[twocolumn,nofootinbib,amsmath,amssymb,aps,prd,balancelastpage,superscriptaddress]{revtex4-1}

\usepackage{color}
\usepackage{xcolor}
\usepackage[active]{srcltx}
\usepackage{amsmath,amsfonts,amssymb,amsthm,amstext,amscd,eucal,srcltx}
\usepackage{epsfig,graphicx,bm}
\usepackage{epstopdf, epsf}
\usepackage{dcolumn}
\usepackage{comment}
\usepackage{hyperref}
\usepackage{tikz}
\usepackage[nointegrals]{wasysym}
\usepackage{braket}
\usepackage{etoolbox}

\usepackage{hyperref}
\usepackage{xr}
\usepackage{xr-hyper}

\newcommand{\ee}{\end{equation}}

\newcommand{\bc}{\begin{center}}
\newcommand{\ec}{\end{center}}

\newcommand{\nn}{\nonumber}

\usepackage{amsmath}
\usepackage{amssymb}
\usepackage{amsthm}
\usepackage{graphicx}
\usepackage{lineno}
\usepackage{float}
\usepackage[all]{xy}
\usepackage{mathrsfs}
\usepackage{tikz-cd}

\usepackage{xcolor}
\usepackage{tikz}
\usetikzlibrary{tqft}
\usetikzlibrary{positioning}

\newcommand{\bea}{\begin{eqnarray}}
\newcommand{\eea}{\end{eqnarray}}

\usepackage{braket}

\newcommand{\ga}{\gamma}

\newcommand{\cH}{\mathcal H}

\newcommand{\cL}{\mathcal L}

\newcommand{\bA}{\mathbb{A}}
\newcommand{\bB}{\mathbb{B}}

\newcommand{\bF}{\mathbb{F}}

\newcommand{\la}{\leftarrow}

\usepackage{graphicx}
\usepackage{epsfig,latexsym}

\usepackage{ulem}
\usepackage{outlines}
\usepackage{enumitem}
\setenumerate[1]{label=\Roman*.}
\setenumerate[2]{label=\Alph*.}
\setenumerate[3]{label=\roman*.}
\setenumerate[4]{label=\alph*.}

\definecolor {darkgreen}{rgb}{0.2,0.7,0.2}

\usepackage{tikz}
\usetikzlibrary{decorations}
\usetikzlibrary{decorations.text}
\usetikzlibrary{arrows}
\tikzset{l/.style={draw=black, line width=1pt}}
\tikzset{s/.style={dashed, draw=black, line width=0.2pt}}

\begin{document}

\newcommand{\nl}{\nonumber\\}

\newcommand{\ans}{ansatz }
\newcommand{\mat}[4]{\left(\begin{array}{cc}{#1}&{#2}\\{#3}&{#4}
\end{array}\right)}
\newcommand{\matr}[9]{\left(\begin{array}{ccc}{#1}&{#2}&{#3}\\
{#4}&{#5}&{#6}\\{#7}&{#8}&{#9}\end{array}\right)}
\newcommand{\matrr}[6]{\left(\begin{array}{cc}{#1}&{#2}\\
{#3}&{#4}\\{#5}&{#6}\end{array}\right)}
\newcommand{\cvb}[3]{#1^{#2}_{#3}}
\def\lsim{\raise0.3ex\hbox{$\;<$\kern-0.75em\raise-1.1ex
e\hbox{$\sim\;$}}}
\def\gsim{\raise0.3ex\hbox{$\;>$\kern-0.75em\raise-1.1ex
\hbox{$\sim\;$}}}
\def\abs#1{\left| #1\right|}
\def\simlt{\mathrel{\lower2.5pt\vbox{\lineskip=0pt\baselineskip=0pt
           \hbox{$<$}\hbox{$\sim$}}}}
\def\simgt{\mathrel{\lower2.5pt\vbox{\lineskip=0pt\baselineskip=0pt
           \hbox{$>$}\hbox{$\sim$}}}}
\def\unity{{\hbox{1\kern-.8mm l}}}
\newcommand{\eps}{\varepsilon}
\def\ep{\epsilon}
\def\ga{\gamma}
\def\Ga{\Gamma}
\def\om{\omega}
\def\omp{{\omega^\prime}}
\def\Om{\Omega}
\def\la{\lambda}
\def\La{\Lambda}
\def\al{\alpha}
\newcommand{\ov}{\overline}
\renewcommand{\to}{\rightarrow}
\renewcommand{\vec}[1]{\mathbf{#1}}
\def\tm{{\widetilde{m}}}
\def\mcirc{{\stackrel{o}{m}}}
\newcommand{\Dm}{\Delta m}
\newcommand{\tanb}{\tan\beta}
\newcommand{\nbar}{\tilde{n}}
\newcommand\PM[1]{\begin{pmatrix}#1\end{pmatrix}}
\newcommand{\up}{\uparrow}
\newcommand{\down}{\downarrow}
\def\omE{\omega_{\rm Ter}}
%

\newcommand{\Dsusy}{{susy \hspace{-9.4pt} \slash}\;}
\newcommand{\DCP}{{CP \hspace{-7.4pt} \slash}\;}
\newcommand{\mc}{\mathcal}
\newcommand{\gr}{\mathbf}
\renewcommand{\to}{\rightarrow}
\newcommand{\gtc}{\mathfrak}
\newcommand{\wh}{\widehat}
\newcommand{\br}{\langle}
\newcommand{\kt}{\rangle}

\newcommand{\Pl}{{\mbox{\tiny Pl}}}
\newcommand{\stat}{{\mbox{\tiny stat}}}
\newcommand{\tot}{{\mbox{\tiny tot}}}
\newcommand{\sys}{{\mbox{\tiny sys}}}
\newcommand{\GW}{{\mbox{\tiny GW}}}
\newcommand{\ny}[1]{\textcolor{blue}{\it{\textbf{ny: #1}}} }
\newcommand{\am}[1]{\textcolor{red}{\it{\textbf{am: #1}}} }

\newcommand{\Hor}{{\mbox{\tiny H}}}
\newcommand{\BH}{{\mbox{\tiny BH}}}
\newcommand{\HL}{{\mbox{\tiny HL}}}
\newcommand{\Bondi}{{\mbox{\tiny Bondi}}}
\newcommand{\DM}{{\mbox{\tiny DM}}}
\newcommand{\Rel}{{\mbox{\tiny Rel}}}
\newcommand{\IGM}{{\mbox{\tiny IGM}}}
\newcommand{\ISM}{{\mbox{\tiny ISM}}}
\newcommand{\CMB}{{\mbox{\tiny CMB}}}
\newcommand{\DE}{{\mbox{\tiny DE}}}
\newcommand{\tidal}{{\mbox{\tiny tidal}}}
\newcommand{\nonspin}{{\mbox{\tiny no-spin}}}
\newcommand{\spinalign}{{\mbox{\tiny spin-aligned}}}
\newcommand{\comp}{{\mbox{\tiny comp}}}


\def\lsim{\mathrel{\mathop  {\hbox{\lower0.5ex\hbox{$\sim$}
\kern-0.8em\lower-0.7ex\hbox{$<$}}}}}
\def\gsim{\mathrel{\mathop  {\hbox{\lower0.5ex\hbox{$\sim$}
\kern-0.8em\lower-0.7ex\hbox{$>$}}}}}

\def\nn{\\  \nonumber}
\def\de{\partial}
\def\brf{{\mathbf f}}
\def\bbf{\bar{\bf f}}
\def\bF{{\bf F}}
\def\bbF{\bar{\bf F}}
\def\bA{{\mathbf A}}
\def\bB{{\mathbf B}}
\def\bG{{\mathbf G}}
\def\bI{{\mathbf I}}
\def\bM{{\mathbf M}}
\def\bY{{\mathbf Y}}
\def\bX{{\mathbf X}}
\def\bS{{\mathbf S}}
\def\bb{{\mathbf b}}
\def\bh{{\mathbf h}}
\def\bg{{\mathbf g}}
\def\bla{{\mathbf \la}}
\def\bmu{\mathbf m }
\def\by{{\mathbf y}}
\def\bmu{\mbox{\boldmath $\mu$} }
\def\bsig{\mbox{\boldmath $\sigma$} }
\def\bunity{{\mathbf 1}}
\def\cA{{\cal A}}
\def\cB{{\cal B}}
\def\cC{{\cal C}}
\def\cD{{\cal D}}
\def\cF{{\cal F}}
\def\cG{{\cal G}}
\def\cH{{\cal H}}
\def\cI{{\cal I}}
\def\cL{{\cal L}}
\def\cN{{\cal N}}
\def\cM{{\cal M}}
\def\cO{{\cal O}}
\def\cR{{\cal R}}
\def\cS{{\cal S}}
\def\cT{{\cal T}}
\def\eV{{\rm eV}}
%

\title{Entropy Production and the Gravitational Origin of the Second Law}

\author{Simone Antonini}
\affiliation{Sapienza University, Piazzale Aldo Moro n.2, Rome, Italy, EU}

\author{Antonino Marcian\`o}
\email{marciano@fudan.edu.cn}
\affiliation{Center for Field Theory and Particle Physics \& Department of Physics, Fudan University, 200433 Shanghai, China}
\affiliation{Laboratori Nazionali di Frascati INFN, Frascati (Rome), Italy, EU}
\affiliation{INFN, sezione Roma Tor Vergata, Rome, I-00133, Italy, EU}

\begin{abstract}
\noindent
We investigate the relation between gravitational dynamics and the second law of thermodynamics in a non-equilibrium framework. Extending Jacobson's thermodynamic derivation of the Einstein equations, we introduce a stochastic geometric flow for the spacetime metric and define entropy production as the ratio between forward and time-reversed trajectories. We show that entropy production is governed by curvature and matter contributions, and that its vanishing selects configurations satisfying the Einstein field equations. Classical general relativity thus emerges as the reversible limit of an underlying stochastic geometro-dynamics, while the second law arises from its non-equilibrium evolution.
\end{abstract}


\maketitle


\vspace{0.3cm}

\section{Introduction} 
\noindent 
The relation between gravitation and thermodynamics has long suggested the existence of a deeper underlying structure. A milestone in this direction was achieved by \cite{Jacobson_1995}, who showed that the Einstein field equations can be derived from the first law of thermodynamics, $\delta Q = T \delta S$, applied to local causal horizons, together with the assumption that entropy is proportional to the horizon area. In this framework, spacetime dynamics emerges as an equation of state, valid in local equilibrium. This result has motivated a broad line of research exploring the thermodynamic interpretation of gravitational dynamics \cite{Padmanabhan_2010}.

This perspective raises a natural and fundamental question: \emph{what is the role of non-equilibrium thermodynamics in gravitational dynamics?} In particular, while Jacobson's construction relies on reversible transformations, physical systems generically evolve through irreversible processes characterized by entropy production. Understanding how such irreversibility is encoded in spacetime dynamics is essential for extending the thermodynamic interpretation of gravity beyond equilibrium. Stochastic formulations of field dynamics have a long history, notably in the context of stochastic quantization \cite{Parisi_Wu_1981, lulli2022stochasticquantizationgeneralrelativity}.

Jacobson's equilibrium construction was later extended to non-equilibrium
spacetime thermodynamics, where an internal entropy-production term is
associated with gravitational dissipation, horizon shear and tidal heating
\cite{Eling_2006,Chirco_Liberati_2010}. This supports the view that
irreversibility is not merely added to gravity, but may be intrinsically
encoded in gravitational degrees of freedom. In parallel, the emergent-gravity
perspective suggests that Einstein gravity itself may arise as a low-energy,
hydrodynamic or collective phenomenon \cite{Barcelo_Liberati_Visser_2001}.
Our approach differs from these works by reversing the usual logic: rather
than deriving gravity from thermodynamic assumptions, we derive entropy
production from a stochastic geometric dynamics, with Einstein gravity
identified as the reversible limit.

In this work, we address this question by introducing a stochastic extension of geometric evolution, based on a Ricci flow supplemented by multiplicative noise. The stochastic dynamics is interpreted as a non-equilibrium relaxation process for the metric, analogous to gradient flows in complex systems. Within this framework, we employ methods from stochastic thermodynamics \cite{Seifert_2012} and express entropy production as the logarithmic ratio between forward and time-reversed trajectories in the space of metrics, following the Onsager-Machlup formalism \cite{Onsager_Machlup_1953}.

This construction leads to three main results. First, we derive an explicit expression for entropy production in terms of geometric quantities, showing that it is governed by curvature and matter contributions. Second, we demonstrate that the condition of vanishing entropy production selects configurations satisfying the Einstein field equations, thereby identifying classical general relativity as the reversible limit of the stochastic dynamics. Third, we analyze the structure of entropy production away from equilibrium and show that it is associated with geometric fluctuations of the metric, providing a natural link between irreversibility and the dynamical evolution of spacetime geometry.

This approach provides a complementary viewpoint to the thermodynamic derivation of gravitational dynamics: rather than starting from thermodynamic principles to obtain the Einstein equations, we derive thermodynamic irreversibility from an underlying stochastic geometro-dynamics. In this sense, the second law of thermodynamics emerges as a consequence of the non-equilibrium relaxation of the gravitational field.\\

\section{Stochastic geometric flow} 
\noindent
We consider a stochastic extension of the Ricci flow for the spacetime metric, originally introduced in a geometric context \cite{Hamilton_1982} and later developed in relation to entropy functionals \cite{Perelman_2002},

\begin{equation}
\label{eq:stochastic_ricci}
\partial_s g_{\mu\nu}(x,s)
=
-2\Big(R_{\mu\nu} - \kappa \,R_{\mu\nu}^{T}\Big)
+ g_{\mu\nu}(x,s)\,\eta(x,s)\,,
\end{equation}
where $s$ is a stochastic (thermodynamic) time parameter, $R_{\mu\nu}$ is the Ricci tensor, $R_{\mu\nu}^{T} = T_{\mu\nu} - \frac{1}{2}T g_{\mu\nu}$ encodes matter contributions, and $\eta$ is a Gaussian noise with zero mean and correlator
\begin{equation}
\langle \eta(x,s)\eta(x',s') \rangle
=
2D\,\frac{\delta^4(x-x')}{\sqrt{-g}}\,\delta(s-s')\,.
\end{equation}

This dynamics can be interpreted as a geometric relaxation process in the space of metrics, driven by curvature and stochastic fluctuations. Functional identities for Gaussian noise, such as the Novikov theorem \cite{Novikov,Furutsu}, play a key role in evaluating stochastic averages.

The stochastic evolution can equivalently be described in terms of a Fokker--Planck equation \cite{Gardiner,Risken}. The induced evolution of geometric quantities under the stochastic flow is discussed in Sec.~\ref{SM:geometry_evolution} of the Appendix. 
The probability of a trajectory $g_{\mu\nu}(s)$ is defined through the Onsager-Machlup functional \cite{Onsager_Machlup_1953},
\begin{equation}
\label{eq:OM_action}
P[g] \propto \exp\left[-S_{\mathrm{OM}}[g]\right]\,,
\end{equation}
where, to leading order in the diffusion coefficient $D$, the action takes the schematic form
\begin{equation}
S_{\mathrm{OM}}[g]
=
\frac{1}{64D}
\int ds\, d^4x \sqrt{-g}
\left[g^{\mu\nu}
\left(\partial_s g_{\mu\nu}
+ 2R_{\mu\nu}
+ \dots
\right)\right]^2\,.
\end{equation}
A derivation of the Onsager--Machlup functional from the stochastic dynamics, 
%
as well as the corresponding Fokker--Planck description and stationary distribution, are discussed in Sec.~\ref{SM:FP} of the Appendix.\\

\section{Path-integral representation} 
\noindent 
The stochastic dynamics admits an equivalent path-integral representation. Introducing a functional delta enforcing the Langevin equation and integrating over the noise, the trajectory probability can be written as
\begin{equation}
P[g] = \int \mathcal{D}\eta\, P[\eta]\, \delta\!\left[\partial_s g_{\mu\nu} + 2R_{\mu\nu} - g_{\mu\nu}\eta \right].
\end{equation}
Using a standard Martin--Siggia--Rose--Janssen--de Dominicis construction \cite{MSR,DeDominicis}, this can be recast as
\begin{equation}
P[g] \propto \int \mathcal{D}g' \exp\left(-S_{\rm MSR}[g,g']\right),
\end{equation}
where $g'_{\mu\nu}$ is an auxiliary response field. Integrating out $g'_{\mu\nu}$ yields the Onsager--Machlup functional,
\begin{eqnarray}
&&S_{\rm OM} \sim \frac{1}{64D} \int ds\, d^4x \{\sqrt{-g}
[g^{\mu\nu}(\partial_s g_{\mu\nu} + 2R_{\mu\nu}+\nonumber\\
&&+\frac{D}{V_{Pl.}}g_{\mu\nu})]^2+..\}\,, 
\end{eqnarray}
with $V_{Pl.}$ Planck size volume, up to terms that do not contribute to entropy production. Further details are provided in Sec.~\ref{SM:OM} of the Appendix.\\

\section{Entropy production} 
\noindent 
Following stochastic thermodynamics \cite{Seifert_2012}, entropy production along a trajectory is defined as the logarithmic ratio between forward and time-reversed path probabilities,
\begin{equation}
\label{eq:entropy_def}
\Delta S
=
\ln \frac{P[g]}{P_R[g]}\,.
\end{equation}
In a fully consistent formulation, the trajectory probability contains both the dynamical Onsager--Machlup contribution and the probability weight induced by the Fokker-Planck evolution. Accordingly, the entropy variation of the Universe can be decomposed as 
$\Delta S_{\rm U} = \Delta S_{\rm OM} + \Delta S_{\rm FP}$, 
where $\Delta S_{\rm OM}$ arises from the stochastic action, while $\Delta S_{\rm FP}$ accounts for the change in probability measure along the flow. Further details concerning the definition of entropy, time-reversal condition and the form of both actions can be found in Secs.~\ref{SM:entropy}-\ref{SM:time_temperature} of the Appendix.

In the regime considered here, either near stationarity or at leading order in the weak-noise expansion, the Fokker--Planck contribution does not modify the antisymmetric part under time reversal that controls irreversibility. The entropy production is therefore governed by the Onsager--Machlup sector, yielding
\begin{equation}
\label{eq:entropy_general}
\Delta S_{\rm OM}
\propto
\left\langle
\int ds \int d^4x \sqrt{-g}\,
g^{\mu\nu}(\partial_s g_{\mu\nu})
\left(
R + \mathcal{O}(D)
\right)
\right\rangle\,.
\end{equation}

Using $\partial_s \sqrt{-g} = \frac{1}{2}\sqrt{-g}\, g^{\mu\nu}\partial_s g_{\mu\nu}$, this expression can be rewritten in geometric form,
\begin{equation}
\label{eq:entropy_compact}
\Delta S_{\rm OM}
=\frac{-1}{4D}
\left\langle
\int ds \int d^4x \,
\partial_s(\sqrt{-g})\, R
\right\rangle
+\mathcal{O}(D)\,.
\end{equation}

This structure makes explicit that entropy production is governed by the coupling between metric flow and curvature. The positivity of entropy production follows from standard fluctuation relations and Jensen's inequality, $\langle e^{-\Delta S}\rangle=1 \Rightarrow \langle \Delta S\rangle \ge 0$, ensuring consistency with the second law --- see 
the relevant sections of the Appendix.\\

\section{Einstein equations as reversible limit} 
\noindent 
The rate of entropy production follows from Eq.\eqref{eq:entropy_compact} as
\begin{equation}
\label{eq:entropy_rate}
\frac{\delta S_{\rm Un.}}{\delta s}
=
\int d^4x \sqrt{-g}\,
\mathcal{F}(g_{\mu\nu},T_{\mu\nu})\,,
\end{equation}
where, to leading order,
$\mathcal{F}
\propto
g^{\mu\nu}
\left(
R_{\mu\nu}
- \kappa\,R_{\mu\nu}^{T}
- \Lambda g_{\mu\nu}
\right)$.

A central result of this work is that the condition of vanishing entropy production,
\begin{equation}
\label{eq:reversible_condition}
\frac{\delta S}{\delta s} = 0\,,
\end{equation}
selects configurations satisfying
$R_{\mu\nu} - \frac{1}{2}R g_{\mu\nu} + \Lambda g_{\mu\nu} = \kappa \,T_{\mu\nu}$,
namely the Einstein field equations.

This establishes that classical general relativity corresponds to the \emph{reversible sector} of the stochastic geometric dynamics, while deviations from Einstein solutions are associated with entropy production and thus with irreversible processes.

Combining the entropy due to the Onsager--Machlup action and the one that is derived from the Fokker-Planck equation, we finally find that 
\begin{eqnarray}\label{V3:entropy-production}
    &&\frac{\delta S_{\rm U}}{\delta s}\propto-\left \langle\int d^4x \frac{\partial_s \sqrt{-g}}{V_{\rm Pl. }} \right \rangle=  \nonumber \\
   && =\left \langle \int d^4x \sqrt{-g} \frac{1}{2V_{\rm Pl.}}g^{\mu\nu}(2(R_{\mu\nu}-kR_{\mu\nu}^T)-\Lambda g_{\mu\nu}\right \rangle\,, \nonumber\\
\end{eqnarray}
where $\Lambda={D}/{V_{\rm Pl.}}$.

Thus, denoting with dot the thermal-time derivative, the Einstein equations (up to term $D^{\frac{1}{2}}$) imply that $\dot{S}=0$, and that $\dot{S}=0$ (and consequently $TdS=\delta Q$), implies the Einstein equations --- performing a saddle point approximation, and requesting the conservation of the stress energy tensor and the stability of the reversible evolution.\\

\section{Irreversibility and geometric flow} 
\noindent 
Away from equilibrium, Eq.~\eqref{eq:entropy_rate} implies
\begin{equation}
\frac{\delta S}{\delta s} \ge 0\,,
\end{equation}
which constrains the admissible geometric evolution. Introducing the local entropy production density 
$\sigma(x,s) \equiv \frac{1}{\sqrt{-g}}\frac{\delta S}{\delta s}$,
one finds, to leading order,
\begin{equation}
\sigma \sim g^{\mu\nu}\partial_s g_{\mu\nu}\, R + \mathcal{O}(D)\,.
\end{equation}

It is important to stress that the total entropy variation receives contributions from both the Fokker--Planck measure and the Onsager--Machlup functional, 
$\Delta S = \Delta S_{\rm FP} + \Delta S_{\rm OM}$. 
In this decomposition, $\Delta S_{\rm FP}$ corresponds to the change of probability density along the stochastic flow and can be interpreted, in the sense of Einstein fluctuation theory, as the entropy variation of the macroscopic (coarse-grained) system. Conversely, $\Delta S_{\rm OM}$ arises from the dynamical path weight and encodes the entropy exchange with the underlying non-equilibrium degrees of freedom, playing the role of an effective environment contribution.

In the near-stationary regime considered here, the antisymmetric part governing irreversibility is dominated by the Onsager--Machlup contribution, while the Fokker--Planck term affects only subleading or boundary contributions.

Using the stochastic flow \eqref{eq:stochastic_ricci}, the Onsager--Machlup sector can be expressed in terms of deviations from the Einstein manifold. Defining 
$X_{\mu\nu} \equiv R_{\mu\nu} - \kappa R_{\mu\nu}^{T} - \Lambda g_{\mu\nu}$, 
one finds
\begin{equation}
\frac{\delta S_{\rm OM}}{\delta s}
\sim
\int d^4x \sqrt{-g}\,
X_{\mu\nu} X^{\mu\nu}
+ \mathcal{O}(D),
\end{equation}
which is manifestly non-negative at leading order.

This shows that entropy production measures the distance from the Einstein sector in configuration space, while the positivity of the total entropy production is ensured by the combined contribution of $\Delta S_{\rm FP}$ and $\Delta S_{\rm OM}$, consistently with Jensen’s inequality and fluctuation relations.

At trace level, projecting onto scalar degrees of freedom, one obtains the behavior
\begin{equation}
\label{eq:entropy_curvature}
\frac{\delta S}{\delta s}
\propto
\int d^4x \sqrt{-g}\,
\left(
R + kT
\right),
\end{equation}
where $T = g^{\mu\nu}T_{\mu\nu}$. The derivation of this projection is detailed in Sec.~\ref{SM:entropy} of the Appendix.

This structure mirrors the standard non-equilibrium form $\dot S=\int J\cdot X$, where the metric flow plays the role of a generalized flux,
$J_{\mu\nu} \sim \partial_s g_{\mu\nu}$,
and curvature defines the thermodynamic force,
\begin{equation}
X^{\mu\nu} \sim R^{\mu\nu} - \kappa R_T^{\mu\nu} - \Lambda g^{\mu\nu}.
\end{equation}
The entropy production therefore measures the distance from the Einstein manifold in configuration space. The second law emerges as a geometric selection principle: stochastic evolution drives the system toward configurations that minimize this distance, identifying Einstein geometries as fixed points of the flow.

As shown in Sec.~\ref{posit} of the Appendix, the expression for the entropy of the universe that appears in Eq.~\eqref{V3:entropy-production} is equivalent to the Smarr formula and gives naturally  rise, for reversible transformation, to the Komar mass.\\

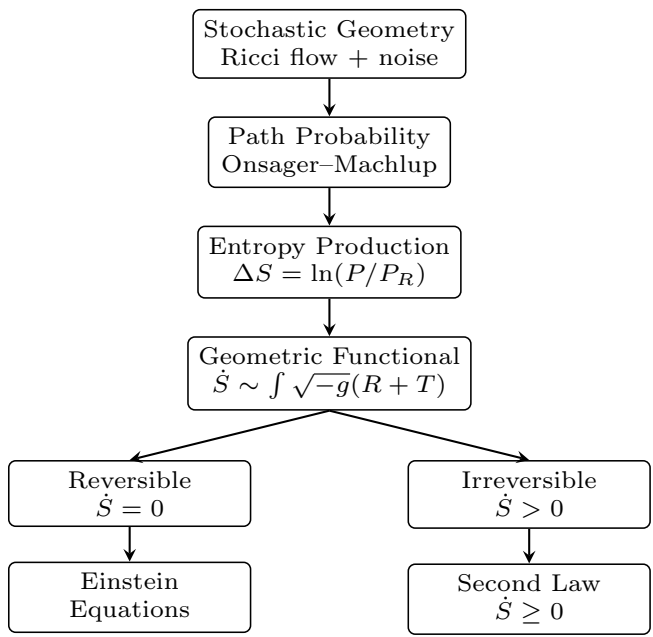
\begin{figure}[t]
\centering
\resizebox{\columnwidth}{!}{%
\begin{tikzpicture}[
    every node/.style={
        draw,
        rounded corners=2pt,
        align=center,
        inner sep=3pt,
        minimum width=2.45cm,
        font=\scriptsize,
        line width=0.45pt
    },
    arrow/.style={->, >=stealth, semithick}
]

\node (stochastic) {Stochastic Geometry\\Ricci flow + noise};
\node (paths) [below=0.38cm of stochastic] {Path Probability\\Onsager--Machlup};
\node (entropy) [below=0.38cm of paths] {Entropy Production\\$\Delta S=\ln(P/P_R)$};
\node (functional) [below=0.38cm of entropy] {Geometric Functional\\$\dot S\sim\int\sqrt{-g}(R+T)$};

\node (einstein) [below left=0.5cm and -0.65cm of functional] {Reversible\\$\dot S=0$};
\node (irreversible) [below right=0.5cm and -0.65cm of functional] {Irreversible\\$\dot S>0$};

\node (GR) [below=0.35cm of einstein] {Einstein\\Equations};
\node (secondlaw) [below=0.35cm of irreversible] {Second Law\\$\dot S\geq0$};

\draw[arrow] (stochastic) -- (paths);
\draw[arrow] (paths) -- (entropy);
\draw[arrow] (entropy) -- (functional);

\draw[arrow] (functional.south) -- (einstein.north);
\draw[arrow] (functional.south) -- (irreversible.north);

\draw[arrow] (einstein) -- (GR);
\draw[arrow] (irreversible) -- (secondlaw);

\end{tikzpicture}
}
\caption{Conceptual structure of the framework. A stochastic geometric flow defines path probabilities for metric configurations. Entropy production, defined as the ratio between forward and time-reversed trajectories, induces a curvature-driven functional. The vanishing of entropy production selects the Einstein equations as the reversible limit, while positive entropy production yields the second law.}
\label{fig:conceptual_flow}
\end{figure}

\section{Discussion}  
\noindent 
The framework developed in this work provides a unified perspective on gravitational dynamics and thermodynamic irreversibility. By formulating a stochastic extension of geometric evolution, we have shown that entropy production can be expressed directly in terms of curvature and matter contributions, and that the Einstein field equations emerge as the condition of vanishing entropy production.  
The relation between the stochastic flow parameter and thermodynamic quantities is further discussed in Sec.~\ref{SM:time_temperature} of the Appendix.

A central implication of this result is that classical general relativity can be interpreted as the \emph{reversible sector} of a more general stochastic geometro-dynamics. In this picture, spacetime configurations evolve in a space of metrics under a flow driven by curvature, while stochastic fluctuations induce deviations from equilibrium. The second law of thermodynamics then arises as a constraint on this evolution, requiring that entropy production be non-negative along admissible trajectories.

This interpretation suggests a natural analogy with gradient flow systems, in which the dynamics is governed by a functional whose stationary points correspond to equilibrium configurations. In the present context, the entropy functional plays this role, with Einstein geometries corresponding to stationary points. The stochastic contributions drive the system toward these configurations while generating irreversible effects away from equilibrium.

Beyond equilibrium, entropy production is controlled by curvature invariants and matter sources, indicating that irreversibility is intrinsically encoded in the geometric structure of spacetime. This provides a geometric realization of the second law, in which entropy production is not imposed as an external principle but emerges from the underlying dynamics of the gravitational field.

An intriguing consequence of this framework is that non-equilibrium evolution may be associated with fluctuations of geometric quantities that are suppressed in the classical limit. While the present analysis focuses on the leading-order stochastic dynamics, it opens the possibility that higher-order effects could encode more refined geometric or topological information. In particular, stochastic fluctuations may provide a mechanism through which spacetime explores nearby configurations in the space of geometries, potentially giving rise to effective coarse-grained descriptions.

It is important to emphasize that the stochastic time parameter $s$ introduced in this work should be interpreted as a thermodynamic or flow parameter, rather than as physical time. This perspective is reminiscent of the thermal time hypothesis, in which time emerges from the statistical state of a system \cite{Rovelli_thermal_time}. In this sense, the evolution described here is analogous to renormalization group flows or gradient descent dynamics, and the emergence of irreversibility should be understood within this extended framework.

Finally, the positivity of entropy in  Eq.~\eqref{V3:entropy-production} provides a insightful instantiation of the arrow of time, which characterizes the evolution of non reversible processes.\\

\section{Conclusions}
\noindent
Several directions for future investigation naturally arise. A first important question concerns the microscopic origin of the stochastic noise and its possible relation to quantum gravitational degrees of freedom. A second direction is the extension of the present formalism to include more general matter sectors and interactions. Finally, it would be interesting to explore possible observational or phenomenological consequences of non-equilibrium geometric dynamics, particularly in regimes where deviations from classical general relativity may become significant.

Overall, the results presented here suggest that thermodynamic irreversibility and gravitational dynamics are deeply intertwined, and that the second law of thermodynamics may be understood as a manifestation of the non-equilibrium evolution of spacetime geometry. In this direction, they resonate with broader ideas relating gravity and thermodynamics \cite{Padmanabhan_2010,Verlinde_2011}.

\appendix 
\section{General remarks and conventions}
\label{SM:conventions}
\noindent 
In this Appendix we summarize the technical steps underlying the main text. Our aim is not to provide a mathematically complete construction of stochastic gravity as a rigorous measure-theoretic theory, but rather to establish the field-theoretic structure required for the entropy-production arguments used in the Letter.

Throughout, $s$ denotes the stochastic or thermodynamic time parameter, distinct from physical time. We consider metric trajectories $g_{\mu\nu}(x,s)$ governed by a stochastic geometric flow. For definiteness, we work in the Stratonovich prescription for multiplicative noise, since it preserves the ordinary chain rule and is therefore the natural choice for geometric stochastic differential equations \cite{Gardiner,Risken}. Whenever needed, the corresponding It\^o form can be obtained by the usual drift renormalization.

In the main text we mostly focus on the vacuum-plus-matter stochastic flow
\begin{equation}
\partial_s g_{\mu\nu}
=
-2\Big(R_{\mu\nu}-\kappa R_{\mu\nu}^{T}\Big)
+ g_{\mu\nu}\eta,
\label{SM:general_flow}
\end{equation}
with
\begin{equation}
R_{\mu\nu}^{T}=T_{\mu\nu}-\frac{1}{2}T g_{\mu\nu},
\qquad
T=g^{\mu\nu}T_{\mu\nu}.
\end{equation}
In several intermediate derivations it is sufficient to set $T_{\mu\nu}=0$ and restore the matter sector at the end by the replacement
\begin{equation}
R_{\mu\nu}\longrightarrow R_{\mu\nu}-\kappa R_{\mu\nu}^{T}.
\end{equation}

We further assume either compact manifolds without boundary or boundary conditions such that total divergences do not contribute. This is sufficient for the local and trace-level results used in the Letter. A more complete treatment with explicit Gibbons--Hawking--York boundary terms would modify only the boundary contribution in the passages related to the calculation of the entropy, but in general not the final form of the entropy. Further details will be provided in the next sections.

\section{Evolution of geometric quantities with multiplicative noise}
\label{SM:geometry_evolution}
\noindent 
Starting from the stochastic Ricci flow
\begin{equation}
\partial_s g_{\mu\nu} = -2R_{\mu\nu} + g_{\mu\nu}\eta,
\label{SM:ricci_flow}
\end{equation}
We may neglect the contributions of the fluctuations provided by matter fields entering the stress energy tensor.\\

We may provide an intuitive argument about the validity of this approximation. Taking into account a field $\phi^a$ that is subject to the stochastic evolution
\begin{eqnarray}\label{SM:Langevin_phi}
    \partial_s \phi^a=F^a+\alpha\xi^a\,,
\end{eqnarray}
we may consider the related stochastic Ricci flow and expanding the Ricci target
\begin{eqnarray}
 &&\partial_s g_{\mu\nu}=-2R_{\mu\nu}+2kR_{\mu\nu}^T(g,\phi)+g_{\mu\nu}\eta=\nonumber\\
 &&=-2R_{\mu\nu}+2kR_{\mu\nu}^T(g,\phi)|_{\phi_0}+2k\frac{\delta R_{\mu\nu}^T}{\delta \phi^a}(\phi^a-\phi^a_0)+\nonumber\\
 &&+O(\alpha^2)+g_{\mu\nu}\eta \,,
\end{eqnarray}
where $\phi_0$ is the solution to the Langevin equation Eq.~\eqref{SM:Langevin_phi} for vanishing noise $\xi^a$ --- in order to explicit the noise contribution from the matter sector  to the geometric gradient flow, we made use of the previous expansion.\\

Thus, the difference $\phi-\phi_0$ is proportional to the amplitude of the stochastic fluctuation $\alpha\xi^a$. If we consider 
$$\frac{\delta R_{\mu\nu}^T}{\delta \phi^a}\alpha\xi^a<\!
\!<g_{\mu\nu}\eta\,,$$ 
which is reasonable for classical objects or in low density states in which vacuum fluctuations dominates over the matter ones, we can neglect contributions due to $\xi^a$.\\

To assess properly the effectiveness of our approach, we have to demonstrate that the Langevin equation that we provided indeed manages to reproduce results of standard general relativity. Neglecting the stochastic fluctuations of the matter content, this turns out to be a completely reasonable approximation for our purposes. Nonetheless, a more complete treatment would involve a system of coupled Langevin equations for both matter fields and the metric tensor.\\

We may now study the induced evolution of the Levi-Civita connection and curvature.

The Christoffel symbols are
\begin{equation}
\Gamma^\lambda_{\mu\nu}
=
\frac{1}{2}g^{\lambda\alpha}
\left(
\partial_\mu g_{\alpha\nu}
+
\partial_\nu g_{\alpha\mu}
-
\partial_\alpha g_{\mu\nu}
\right).
\end{equation}
Differentiating with respect to $s$ gives
\begin{eqnarray}
\partial_s\Gamma^\lambda_{\mu\nu}
=&&
\frac{1}{2}\,\partial_s g^{\lambda\alpha}
\left(
\partial_\mu g_{\alpha\nu}
+
\partial_\nu g_{\alpha\mu}
-
\partial_\alpha g_{\mu\nu}
\right)+  \\
&&+
\frac{1}{2}g^{\lambda\alpha}
\left(
\partial_\mu \partial_s g_{\alpha\nu}
+
\partial_\nu \partial_s g_{\alpha\mu}
-
\partial_\alpha \partial_s g_{\mu\nu}
\right). \nonumber
\end{eqnarray}

Working in a locally inertial frame at a given point, where $\partial_\rho g_{\mu\nu}=0$, the first term vanishes at that point and one finds
\begin{equation}
\partial_s\Gamma^\lambda_{\mu\nu}
=
\frac{1}{2}g^{\lambda\alpha}
\left(
\partial_\mu \partial_s g_{\alpha\nu}
+
\partial_\nu \partial_s g_{\alpha\mu}
-
\partial_\alpha \partial_s g_{\mu\nu}
\right).
\end{equation}
Substituting Eq.~\eqref{SM:ricci_flow}, we can write
\begin{align}
\partial_s\Gamma^\lambda_{\mu\nu}
&=
-g^{\lambda\alpha}
\left(
\partial_\mu R_{\alpha\nu}
+
\partial_\nu R_{\alpha\mu}
-
\partial_\alpha R_{\mu\nu}
\right)+
\nonumber\\
&\quad
+
\frac{1}{2}g^{\lambda\alpha}
\left(
\partial_\mu(g_{\alpha\nu}\eta)
+
\partial_\nu(g_{\alpha\mu}\eta)
-
\partial_\alpha(g_{\mu\nu}\eta)
\right).
\end{align}
Restoring covariance, we find 
\begin{align}
\partial_s\Gamma^\lambda_{\mu\nu}
&=
-g^{\lambda\alpha}
\left(
\nabla_\mu R_{\alpha\nu}
+
\nabla_\nu R_{\alpha\mu}
-
\nabla_\alpha R_{\mu\nu}
\right)
+\mathcal{N}^\lambda_{\mu\nu},
\label{SM:gamma_evolution}
\end{align}
with noise contribution
\begin{equation}
\mathcal{N}^\lambda_{\mu\nu}
=
\frac{1}{2}
\left(
\delta^\lambda_{\nu}\nabla_\mu\eta
+
\delta^\lambda_{\mu}\nabla_\nu\eta
-
g_{\mu\nu}\nabla^\lambda\eta
\right).
\end{equation}

The Riemann tensor evolves as
\begin{equation}
\partial_s R^\lambda{}_{\mu\nu\rho}
=
\nabla_\nu (\partial_s\Gamma^\lambda_{\mu\rho})
-
\nabla_\rho (\partial_s\Gamma^\lambda_{\mu\nu}),
\label{SM:riemann_evolution}
\end{equation}
which makes explicit that curvature evolution is controlled by Ricci flow plus derivatives of the noise.

Contracting indices yields the Ricci tensor evolution,
\begin{eqnarray}
\partial_s R_{\mu\nu}
=
\Delta_L R_{\mu\nu}
&&+
\text{quadratic curvature terms} \nonumber \\
&&+
\text{noise-gradient terms},
\end{eqnarray}
where $\Delta_L$ denotes the Lichnerowicz Laplacian. At trace level one recovers
\begin{equation}
\partial_s R
=
\Delta R + 2R_{\mu\nu}R^{\mu\nu} + \text{noise corrections}.
\label{SM:R_evolution_schematic}
\end{equation}
Equation \eqref{SM:R_evolution_schematic} is the only ingredient needed for the entropy-density evolution used below.

\section{Fokker--Planck equation and stationary distribution}
\label{SM:FP}
\noindent 
The stochastic flow \eqref{SM:ricci_flow} may be written in Langevin form \cite{Gardiner,Risken}
\begin{equation}
d g_{\mu\nu} = -2R_{\mu\nu}\, ds + g_{\mu\nu}\circ dW,
\label{SM:langevin}
\end{equation}
where $\circ$ denotes Stratonovich multiplication and
\begin{equation}
\langle dW(x,s)dW(x',s')\rangle
=
2D\,\frac{\delta^4(x-x')}{\sqrt{-g}}\,\delta(s-s')\,ds\,ds'.
\end{equation}

The corresponding Fokker--Planck equation for the probability functional $P[g,s]$ acquires the form, using the Stratonovich prescription,
\begin{equation}
\partial_s P[g,s]
=
-\frac{\delta}{\delta g_{\mu\nu}}
\left(F_{\mu\nu}P\right)
+
D\frac{\delta}{\delta g_{\mu\nu}}
\left(
g_{\mu\nu}
\frac{\delta g_{\alpha\beta}(x) P}{\delta g_{\alpha\beta}(x)}
\right),
\label{SM:FP_equation}
\end{equation}
where
\begin{equation}
F'_{\mu\nu}=-2R_{\mu\nu}.
\end{equation}

A more covariant treatment would include the DeWitt supermetric and associated functional measure factors. Since the Letter only uses the trace-level stationary structure, the simplified form \eqref{SM:FP_equation} is sufficient. In particular, it correctly captures the competition between deterministic Ricci drift and multiplicative diffusion.

The stationary condition $\partial_s P=0$ implies
\begin{equation}
\frac{\delta}{\delta g_{\mu\nu}}
\left(F_{\mu\nu}P\right)
=
D\frac{\delta}{\delta g_{\mu\nu}}
\left((Bg_{\mu\nu}) P+
g_{\mu\nu}g_{\alpha\beta}
\frac{\delta P}{\delta g_{\alpha\beta}}
\right)\,,
\end{equation}
with {$B^{\mu\nu}\,_{\rho\sigma}=\mathbf{1}^{\mu\nu}\,_{\rho\sigma}$ (with $\mathbf{1}^{cd}_{ab}=\frac{1}{2}(\delta^c_a\delta^d_b+\delta^c_b\delta^d_a)$.\\

At leading order this is solved by
\begin{equation}
P[g]\propto
\exp\left[
-\frac{1}{2D}S_{\rm EH}[g]
+\frac{B}{2}\int d^4x\,\sqrt{-g}\,\delta(0)
\right],
\label{SM:stationary}
\end{equation}
where $S_{\rm EH}$ is the Einstein--Hilbert action. The $\delta(0)=\delta^4(0)/\sqrt{-g}$ term is the standard local contribution associated with coincident functional contractions and may be interpreted as a volume-density counterterm. In a local inertial frame we can indeed consider it as the inverse of the Planck volume $V_{\rm Pl}$. Since the expression is covariant, the identification holds in any reference frame. This does not affect the antisymmetric time-reversal sector that controls entropy production.

Thus the stationary distribution is controlled by the same functional that defines classical general relativity. This is the first signal that Einstein configurations play the role of equilibrium configurations of the stochastic dynamics.

\section{Noise distribution}
\label{SM:noise}
\noindent 
The Gaussian noise is taken to have probability distribution
\begin{equation}
P[\eta]\propto
\exp\left[
-\frac{1}{4D}\int d^4x\,ds\,\sqrt{-g}\,\eta^2
\right],
\label{SM:noise_dist}
\end{equation}
which yields
\begin{equation}
\langle \eta(x,s)\eta(x',s')\rangle
=
2D\,\frac{\delta^4(x-x')}{\sqrt{-g}}\,\delta(s-s').
\label{SM:noise_corr}
\end{equation}

By the Wick's theorem, higher connected correlators vanish identically for the Gaussian measure. This is precisely why the Novikov identity applies and why the stochastic corrections used in the Letter close at two-point level. In the weak-noise regime, higher-order nonlinear corrections arise only through the metric dependence of the multiplicative coupling, not from non-Gaussian noise cumulants.\\

It is worth to notice that, in the distribution, the $\sqrt{-g}$ depends on the noise --- this is useful in the calculation of the path integral action. This does not affect the cumulants, since the contribution are of the order $D^n$ with $n>1$.\\

Since we will see that $D\sim \Lambda V_{\rm Pl}$ (with $\Lambda$ cosmological constant), we can neglect higher order contributions.

\section{Path integral and Onsager--Machlup functional}
\label{SM:OM}
\noindent 
The probability of a metric trajectory may be written as
\begin{equation}
P[g]
=
\int D\eta\,P[\eta]\,
\delta\!\left[
\partial_s g_{\mu\nu}+2R_{\mu\nu}-g_{\mu\nu}\eta
\right]\det[\frac{\delta \mathcal{E}_{\mu\nu}}{\delta g_{\alpha\beta}}].
\label{SM:pathprob}
\end{equation}
with
\begin{equation}
\mathcal{E}_{\mu\nu}
\equiv
\partial_s g_{\mu\nu}+2R_{\mu\nu}-g_{\mu\nu}\eta,
\end{equation}
Using a functional Fourier representation of the delta functional, we recover
\begin{equation}
\delta[\mathcal{E}_{\mu\nu}]
\propto
\int Dg'_{\mu\nu}\,
\exp\left[
-\int ds\,d^4x\,\sqrt{-g}\,
g'^{\mu\nu}\mathcal{E}_{\mu\nu}
\right],
\end{equation}
one obtains the Martin--Siggia--Rose--Janssen--de Dominicis functional representation \cite{MSR,DeDominicis}
\begin{equation}
P[g]\propto
\int Dg'\,\exp\left(-S_{\rm MSR}[g,g']\right).
\end{equation}

Performing the Gaussian integral over $\eta$ gives
\begin{align}
S_{\rm MSR}[g,g']
&\!=\!\!
\int\! ds\,d^4x\,\sqrt{-g}\,
g'^{\mu\nu}\!\left(\partial_s g_{\mu\nu}\!+
\!2R_{\mu\nu}\!+\!BD\frac{g_{\mu\nu}}{V_{\rm Pl}}\right)
\nonumber\\
&\quad
-
D\int ds\,d^4x\,\sqrt{-g}\,
\left(g'^{\mu\nu}g_{\mu\nu}\right)^2
+\cdots,
\end{align}
where the omitted terms arise from Jacobians, measure factors, and coincident-point regularization. These terms are even under time reversal and do not contribute to the antisymmetric sector of the action that defines entropy production.

Integrating out the response field $g'_{\mu\nu}$ at saddle-point level yields the Onsager--Machlup functional \cite{Onsager_Machlup_1953}
\begin{equation}
S_{\rm OM}
=
\int \frac{ ds\,d^4x\,}{64D}\sqrt{-g}\!
\left[
g^{\mu\nu}\!
\left(
\partial_s g_{\mu\nu}+2R_{\mu\nu} +B'D\frac{g_{\mu\nu}}{V_{\rm Pl}}
\right)
\right]^2\!\!,
\label{SM:OM_action}
\end{equation}
with $B'$ constant term.\\

The key point for the Letter is that under time reversal $s\to -s$, the terms even in $\partial_s g_{\mu\nu}$ cancel in the ratio $P[g]/P_R[g]$, while the cross-term linear in $\partial_s g_{\mu\nu}$ survives and produces the entropy-production functional.

\section{Novikov theorem and stochastic contractions}
\label{SM:novikov}
\noindent 
For Gaussian noise, the Novikov identity \cite{Novikov,Furutsu} reads
\begin{equation}
\langle \eta(x,s)F[g]\rangle
=
2D
\left\langle
\frac{\delta F[g]}{\delta \eta(x,s)}
\right\rangle.
\label{SM:novikov_identity}
\end{equation}
Applying this to the metric gives
\begin{equation}
\langle \eta\,g_{\mu\nu}\rangle
=
2D
\left\langle
\frac{\delta g_{\mu\nu}}{\delta\eta}
\right\rangle.
\end{equation}
In the Stratonovich prescription, causality implies \cite{Fox}\cite{Sancho}:
\begin{equation}
\frac{\delta g_{\mu\nu}(x,s)}{\delta\eta(x',s')}
=
\frac{1}{2}
g_{\mu\nu}(x,s)\,
\frac{\delta^4(x-x')}{\sqrt{-g}}\,
\delta(s-s')
+\cdots,
\end{equation}
hence
\begin{equation}\label{SM:novikov con g}
\langle \eta\,g_{\mu\nu}\rangle
=
D\,g_{\mu\nu} \frac{1}{V_{\rm Pl}} 1+O(D^2).
\end{equation}
This relation is sufficient for evaluating the leading stochastic correction to the entropy-production density.

\section{Entropy production: full structure and positivity}
\label{SM:entropy}
\noindent 
Entropy production is defined as
\begin{equation}
\Delta S
=
\ln\frac{P[g]}{P_R[g]}.
\label{SM:entropy_def}
\end{equation}

In a complete stochastic formulation, the trajectory probability contains both:
\begin{itemize}
\item the Onsager--Machlup dynamical contribution,
\item the Fokker--Planck weight associated with the evolving probability measure.
\end{itemize}
In particular we have that: 
\begin{eqnarray}
P[g]=P_0(g_0)P_{\rm OM}\{g_0|g\}\,,\\
P_R[g]=P(g_f)P_{\rm R}(g(s))\,,
\end{eqnarray}
where $P_{\rm OM}\{g_0|g\}$ denotes the probability of a transition (calculated with the path integral) and $P_{\rm R}(g(s))$ represents the same probability calculated under time reversal symmetry, which means $g'(s)=g(t_f-s)$ for $0\leq s \leq t_f$.\\

$P[g_0]$ and $P[g_f]$ are calculated from the FP equation since we suppose to start and end up at stationary states \footnote{The FP equation provides a transition probability. For stationary processes we find that $P[g_0]=\int Dg_i P_{FP}[g_0|g_i]P(g_i)=P_{FP}[g_0|g_i]$, with $P[g_i]$ generic initial probability and $P_{FP}[g_0|g_i]$ stationary solution of the FP equation.}.
\\

Accordingly,
\begin{equation}
\Delta S = \Delta S_{\rm OM} + \Delta S_{\rm FP}.
\end{equation}

It is immediate to recognize that the normalisation of the FP distribution cancels out. In fact, using a change of variable we have that 
\begin{eqnarray}
    \int Dg(t_0)\exp{-\frac{S[g({t_0})]}{2D}}=\int Dg(t_f)\exp{-\frac{S[g({t_f})]}{2D}}\,.
\end{eqnarray}
A bit more involved is the demonstration that the normalization drops out from the path integral expression. For simple cases one can show, adopting the Stratonovich prescription, that it factors out in the ratio between the path probability and its time reversal \cite{Entropy}. More in general, within the Crooks \cite{Crooks} approach, one can write the normalization as a free energy, finally factored out while considering the ratio.\\

We will generally assume that in the ratio the normalizations cancel out and we will neglect its contribution.

\subsection{Onsager--Machlup contribution}
\noindent 
Using the action \eqref{SM:OM_action}, the antisymmetric part under time reversal yields
\begin{equation}
\Delta S_{\rm OM}
\propto\textcolor{blue}{-}
\int ds\,d^4x\,\sqrt{-g}\,
g^{\mu\nu}\partial_s g_{\mu\nu}\left( \,2R +
\frac{D}{V_{\rm Pl}}N\right)\,, 
\label{SM:entropy_from_OM}
\end{equation}
with N constant. Then, further manipulating, one finds

\begin{eqnarray}
 &&\Delta S_{OM}=-\frac{1}{8D}\int ds d^4x \partial_s\sqrt{-g}(2R+N\frac{D}{V_{Pl}})=\nonumber\\
 &&=-\frac{1}{2D}\int d^4x \int ds \frac{\partial \sqrt{-g}}{\partial s}\frac{\partial (S_{EH}+\int d^4y\frac{N'D}{V_{\rm Pl}}\sqrt{-g})}{\partial \sqrt{-g}}=\nonumber\\
 &&=-\frac{1}{2D}\Delta S_{EH}-\int d^4x \Delta \sqrt{-g}\frac{N'}{2V_{\rm Pl}}  \,, 
\end{eqnarray}
with $S_{EH}=\int d^4y \sqrt{-g}R$.\\

In the previous formula we used the following two identities:
\begin{eqnarray}
    &&\frac{\delta S_{EH}}{\delta \sqrt{-g}}=\frac{\delta g^{\mu\nu}}{\delta \sqrt{-g}}\frac{\delta S_{EH}}{\delta g^{\mu\nu}}=\frac{-g^{\mu\nu}}{2\sqrt{-g}}(R_{\mu\nu}-\frac{1}{2}g_{\mu\nu}R)=\frac{R}{2}\,,\nonumber\\
    &&1=\frac{\delta \sqrt{-g}}{\delta \sqrt{-g}}=\frac{\delta g^{\mu\nu}}{\delta \sqrt{-g}}\frac{\delta \sqrt{-g}}{\delta g^{\mu\nu}}\xrightarrow[]{}\frac{\delta g^{\mu\nu}}{\delta \sqrt{-g}}=-\frac{g^{\mu\nu}}{2\sqrt{-g}}\,.
\end{eqnarray}
It is now clear that it will be straightforward to generalize this result to a non vacuum scenario, and to a case in which boundary terms are taken into account. In fact, their contribution will just change the action $S_{EH}$, anyway vanishing in the calculation of the entropy of the Universe.

\subsection{Fokker--Planck contribution}
\noindent 
The Fokker--Planck term corresponds to the ratio of probability densities,
\begin{equation}
\Delta S_{\rm FP}
=
-\ln P[g_f] + \ln P[g_i].
\end{equation}
Using the stationary distribution \eqref{SM:stationary}, this contributes terms proportional to
\begin{equation}
\Delta S_{\rm FP} \sim \frac{1}{2D}\Delta S_{\rm EH}  -C\int d^4x \frac{\Delta \sqrt{-g}}{V_{\rm Pl}}.
\end{equation}

In regimes close to stationarity or at leading order in $D$, this contribution does not affect the antisymmetric part controlling entropy production, and can be consistently absorbed into boundary terms or normalization factors.

\subsection{Geometric entropy production}
\noindent 
Combining the two contributions and retaining leading-order antisymmetric terms yields for the entropy production in the Universe 
\begin{equation}\label{SM:entropia e volume}
\langle \Delta S_{\rm U}\rangle
\propto
-\Big \langle \int ds\,d^4x\,
\Delta(\sqrt{-g} \,)\, \frac{1}{V_{\rm Pl}} \Big \rangle\,.
\end{equation}

As we can see, the terms related to the drift disappear. It is then straightforward to show that this is a property of the Langevin equations endowed with a drift term that can be written as a derivative of another quantity.\\

We then obtain that
\begin{eqnarray}
    &&\Big \langle\frac{\delta S_{\rm U}}{\delta s}\Big \rangle \propto \Big \langle \int d^4x \sqrt{-g}(R-2\eta(x,s)\Big \rangle =\nonumber\\
    &&=\Big \langle \int d^4x \sqrt{-g}g^{\mu\nu}(R_{\mu\nu}-g_{\mu\nu}\Lambda)\Big \rangle\,,
\end{eqnarray}
where we used the results cited in Eq.~\eqref{SM:novikov con g} and defined $\Lambda\propto \frac{D}{V_{\rm Pl}}$.\\

Therefore, since we have that $D\sim \Lambda V_{\rm Pl}$, we can perform a saddle point approximation and neglect the $\langle\, \cdot\, \rangle$, hence obtaining
\begin{eqnarray}
   &&\Big \langle\frac{\delta S_{\rm U}}{\delta s}\Big \rangle=\Big \langle A\int d^4x \sqrt{-g}g^{\mu\nu}(R_{\mu\nu}-g_{\mu\nu}\Lambda)\Big \rangle=\nonumber\\
   &&=A'\int d^4x \sqrt{-g}g^{\mu\nu}(R_{\mu\nu}-g_{\mu\nu}\Lambda)\,.
\end{eqnarray}
Inserting now also matter within the form of the Ricci target we obtain:
\begin{eqnarray}\label{SM:entropy production con Ricci target}
    \Big \langle\frac{\delta S_{\rm U}}{\delta s}\Big \rangle=A'\int d^4x \sqrt{-g}g^{\mu\nu}(R_{\mu\nu}-R_{\mu\nu}^T-g_{\mu\nu}\Lambda)
\end{eqnarray}

We can moreover rewrite the formula for the entropy of the Universe in Eq.~\eqref{SM:entropia e volume} as it follows:\\
\begin{eqnarray}
    &&\langle\delta S_{\rm U}\rangle\propto-\Big\langle\frac{1}{V_{\rm Pl}} \delta V \Big\rangle\,, 
    \end{eqnarray}
with $\delta V$ defined as
    \begin{eqnarray}
    \delta V=\int d^4x \delta(\sqrt{-g})=\int d^4x \sqrt{-g}\, \delta \ln \sqrt{-g}\,.
\end{eqnarray}\\
The logarithmic formula can be interpreted as the gravitational analogous of Gibbs' formula for the entropy.

\subsection{Local entropy production and positivity}
\noindent 
Defining
\begin{equation}
S_{\rm OM} = \int d^4x\,\sqrt{-g}\,\sigma(x,s),
\end{equation}
one finds
\begin{equation}
\partial_s \sigma
=
\frac{1}{4D}\Delta R
+
\frac{1}{2D}R_{\mu\nu}R^{\mu\nu}
-
\frac{1}{2}
\left\langle
\frac{\delta R}{\delta\eta}
\right\rangle.
\end{equation}

Near the Einstein sector it holds 
\begin{equation}
X_{\mu\nu} =
R_{\mu\nu}-\kappa R_{\mu\nu}^T-\Lambda g_{\mu\nu},
\end{equation}
and the stochastic flow then implies
\begin{equation}
\partial_s g_{\mu\nu} \sim -2X_{\mu\nu}.
\end{equation}
Hence,
\begin{equation}
\frac{\delta S_{\rm OM}}{\delta s}
\sim
\int d^4x\,\sqrt{-g}\,
X_{\mu\nu}X^{\mu\nu}
+\mathcal{O}(D).
\label{SM:positive_form}
\end{equation}

This provides an explicit quadratic form ensuring positivity at leading order.


\subsection{Jensen inequality and fluctuation relation}
\noindent
Independently of the approximation, entropy production satisfies the exact identity
\begin{equation}
\left\langle e^{-\Delta S} \right\rangle = 1,
\end{equation}
which implies, by Jensen's inequality,
\begin{equation}
\langle \Delta S \rangle \ge 0.
\end{equation}

Thus positivity is guaranteed both:
\begin{itemize}
\item structurally (quadratic form near Einstein solutions); 
\item statistically (fluctuation theorem).
\end{itemize}

\section{Time--temperature relation}
\label{SM:time_temperature}
\noindent 
The stochastic parameter $s$ should be interpreted as a thermodynamic flow parameter, not as physical time. A heuristic relation to temperature follows from the energy-time uncertainty relation,
\begin{equation}
\Delta t\sim\frac{1}{\Delta E}\sim\frac{1}{T},
\end{equation}
hence
\begin{equation}
\Delta t\propto \frac{1}{T}.
\label{SM:thermal_time}
\end{equation}

The latter relation is relevant for the thermodynamical interpretation that we will provide in the following section.\\

In order to calculate the thermodynamical quantities, including heat and entropy, we will need to take into account a time long enough to generate a sizable heat and entropy flow.\\

Calculating the average in time of the vacuum fluctuations that generate those quantities, we obtain
\begin{eqnarray}
    \langle\Delta t\rangle\sim\Big\langle\frac{
    1
    }{\Delta E}\Big\rangle\sim
    \frac{1}{T}\,,
\end{eqnarray}
with $\langle\Delta t\rangle$ mean value of the time-extensions of the vacuum fluctuations --- thus $\langle F\rangle=\sum_{p_i} g(p_i) F_i$, with $g(p_i)$ probability to create a particle $p_i$ --- and $T$ temperature of the system.\\

We stress that Eq.~\eqref{SM:thermal_time} is not used as a dynamical input in the derivation of the main results. Rather, it provides an interpretive bridge between the stochastic flow parameter and thermodynamic time. 

\section{Recovering Jacobson's formula}
\noindent
Requiring the transformation to be reversible ($\dot{S}=0$) implies, through Eq.~\eqref{SM:entropy production con Ricci target}, the validity of the equations of motion for the trace. If we consider a transformation that remains reversible, regardless the size of the volume, we may then neglect the integral --- while more complex processes would require more complicated procedures.\\

The condition for reversibility is in general 
\begin{equation}\label{SM:Field equation con A}
    G_{\mu\nu}+\Lambda g_{\mu\nu}-kT_{\mu\nu}=A_{\mu\nu}\,,
\end{equation}
with $A_{\mu\nu}$ traceless tensor.\\

Thus, if the transformation is reversible, we find 
\begin{equation}
\partial_sg_{\mu\nu}=2(A_{\mu\nu}-\Lambda g_{\mu\nu})+g_{\mu\nu}\eta\,.
\end{equation}
In order to find the exact form of $A_{\mu\nu}$, it is reasonable to suppose that the reversible transformation is at least as stable (and as simple) as the initial Ricci Flow equation, and therefore it does not contain derivative of the metric higher than the second order and it is, at maximum, linear in the second order derivatives.\\

Furthermore, the conservation of the stress-energy tensor implies through Eq.~\eqref{SM:Field equation con A} that $A_{\mu\nu}$ is divergence-less.\\

Combining these two observations  with the property of symmetry of $A_{\mu\nu}$, we can apply the Lovelock theorem \cite{Lovelock} and find that $A_{\mu\nu}$ has the form
\begin{equation}  A_{\mu\nu}=AG_{\mu\nu}+Bg_{\mu\nu}\,,
\end{equation}
with $A$ and $B$ constants.\\

The traceless condition then provides
\begin{equation}    g^{\mu\nu}A_{\mu\nu}=Ag^{\mu\nu}G_{\mu\nu}+4B=-AR+4B=0\,.
\end{equation}
Since $R$ is not a constant, the only solution to the previous equation is $A=B=0$.\\

Thus, if we require that the transformation is reversible ($\delta S_{\rm U}=0$) --- together with the conservation of the energy and the requirement related to the stability --- the Einstein equations must hold.\\

On the other side, the reversibility requirement implies that, for the system, the Clausius equation holds ($T\delta S_{\rm S}=\delta Q$).\\

Finally, we have shown the validity of Jacobson's formulation, only relying on arguments of stochastic thermodynamics, and without using any specific {\it ansatz} about the area-law for the entropy.

\section{Recovering the Komar mass and the Smarr formula}\label{Komar mass}
\noindent
For simplicity, we have hitherto worked using a boundary-less manifold. This is a non-necessary hypothesis, since we can simply add the Gibbons-Hawking-York (GHY) boundary term \cite{GH,York} to the action and obtain the same result. After a simple calculation, within the same framework and approach used previously, we obtain that the entropy production has the same form as in Eq.~\eqref{SM:entropy production con Ricci target}.\\

We may show now that for a reversible transformation respecting the aforementioned hypotheses (i.e., transformations in which the Einstein equations hold), in a space with a timelike Killing vector ($k^{\mu}$), we can recast the metric in the form
\begin{equation}
g_{\mu\nu}=-u_{\mu}u_{\nu}+h_{\mu\nu}\,,
\end{equation}
with $u^{\mu}$ unit timelike vector. Hence, we can write $k^\mu=u^{\mu}N$ --- for simplicity, we have assumed a vanishing shift vector, i.e. $\beta^i=0$, but in what follows we will consider the more general case  $\beta^i\neq0$ and observe that the results remain unchanged.\\

Denoting now with $K_{ij}$ the extrinsic curvature tensor and with $R^{(3)}$ the 3D Ricci scalar, we can recast the scalar Gauss equation \cite{Gurgolon} as
\begin{eqnarray}\label{SM:scalar Gauss}
    R=-2R_{\mu\nu}u^\mu u^\nu+R^{(3)}+K^2-K_{ij}K^{ij}\,.
\end{eqnarray}
Together with the Hamiltonian constraint \cite{Gurgolon} (neglecting the cosmological constant)
\begin{eqnarray}\label{SM:scalar constraint}
    R^{(3)}+K^2-K_{ij}K^{ij}=2k\rho\,,
\end{eqnarray}
we obtain
\begin{eqnarray}
    R=-2R_{\mu\nu}u^\mu u^\nu+2k\rho\,,
\end{eqnarray}
with $\rho=T_{\mu\nu}u^\mu u^\nu$ energy density.\\

Inserting the latter relation inside Eq.~\eqref{SM:entropy production con Ricci target} --- neglecting again the cosmological constant term --- and writing $g^{\mu\nu}T_{\mu\nu}=-\rho+S$, we obtain
\begin{eqnarray}
    &&\frac{\delta S_{uni}}{\delta s}=\int d^4x 2A'\sqrt{-g}[u^{\mu}u^{\nu}(-2R_{\mu\nu})+2k\rho-k\rho+kS]\nonumber\\
    &&=\int d^4x 2A'\sqrt{-g}[u^{\mu}u^{\nu}(-2R_{\mu\nu})+k\rho+kS]\nonumber=\\
    &&=\int d^4x 4A'\sqrt{-g}[u^{\mu}u^{\nu}(-R_{\mu\nu})+\frac{k}{2}(\rho+S)]\nonumber=0\,.
\end{eqnarray}
Decomposing $ \sqrt{-g}=N\sqrt{h}$, we can then recast the latter relation as 
\begin{equation}\label{SM:Komar}
    \frac{2}{k}\int d^3x \sqrt{h}R_{\mu\nu}k^{\mu}u^{\nu}=2\int d^3x \sqrt{h}k^{\mu}u^{\nu}(T_{\mu\nu}-\frac{1}{2}Tg_{\mu\nu})\,,
\end{equation}
where we exploited a timelike Killing vector while performing the integral over the time coordinate, and the fact --- since $u^\mu u^\nu g_{\mu\nu}=-1$ --- that
\begin{eqnarray}
    u^{\mu}u^{\nu}[T_{\mu\nu}-\frac{1}{2}g_{\mu\nu}T]
    =T_{00}+\frac{1}{2}(-\rho+S)=\frac{1}{2}(\rho+S)\,.
\end{eqnarray}
The term on the left hand-side of Eq.~\eqref{SM:Komar} is precisely the volume-integral definition of the Komar mass \cite{Wald}. It has been demonstrated --- see e.g. \cite{Wald} --- that this relation is equivalent to the Smarr formula for black holes \cite{Smarr}. Therefore, this finally proves that, as expected, the Smarr formula is just a particular case of the relation $T\delta S=\delta Q$.\\

If the shift vector $\beta^i\neq 0$, we obtain an additional term --- since the timelike Killing vector $k^\mu=u^\mu N+\beta^\nu$ \cite{Gurgolon} --- namely
\begin{eqnarray}
    -2R_{\mu\nu}u^\mu u^\nu N=-2R_{\mu\nu}u^\nu k^\mu+2R_{\mu\nu}u^\nu \beta^\mu\,.
\end{eqnarray}
If the transformation would have been assumed to be reversible  --- i.e. $R_{\mu\nu}=T_{\mu\nu}-\frac{1}{2}g_{\mu\nu}T$ --- because of $g_{\mu\nu}u^\mu \beta^\nu=u_\nu \beta^\nu=0$ \cite{Gurgolon}, we would have obtained
\begin{eqnarray}
    &&-2R_{\mu\nu}u^\mu u^\nu N=-2R_{\mu\nu}u^\nu k^\mu+2T_{\mu\nu}u^\nu \beta^\mu=\nonumber\\
    &&=-2R_{\mu\nu}u^\nu k^\mu-2p_\mu\beta^\mu\,,
\end{eqnarray}
with $-p_\nu\beta^\nu=T_{\mu\beta}u^\mu h^{\beta}_\nu\beta^\nu=T_{\mu\beta}u^\mu(g^\beta_\nu+u^\beta u_\nu)\beta^\nu=T_{\mu\nu}u^\mu \beta^\nu$, as defined in \cite{Gurgolon}.\\

In other words, this procedure would have provided the term $-2p_\mu \beta^\mu$ that is present inside the definition of the Komar mass --- see e.g. Ref.~\cite{Gurgolon}.\\

More in general, we can rewrite the entropy production --- neglecting again the cosmological constant --- according to
\begin{eqnarray}
    \frac{\delta S}{\delta s}\propto \Big \langle\int d^4x \sqrt{-g}\, [R-\rho+P_{tot}]\Big \rangle\,,
\end{eqnarray}
where we have used the fact that the trace of the stress-energy tensor can be considered (for a shear-less fluid such as a perfect gas) as $T=-\rho+P_{tot}$ ---  where $\rho$ denotes the energy density and $P_{tot}$ the total pressure generated by the matter content.\\

The relation between $T$, $\rho$ and $P_{tot}$ presents a striking resemblance with the first principle of thermodynamics, since
\begin{eqnarray}
    &&\rho dV=-TdV+P_{tot}dV\quad \quad \text{(Trace of $T_{\mu\nu}$)}\\
    &&dU=\delta Q-PdV\, \quad\qquad \quad\text{(First principle)}
\end{eqnarray}
where the signs for the work are different due to a different sign convention in the Smarr formula \cite{Smarr,Wald}. This brings therefore tom the identification 
\begin{eqnarray}
    -TdV=\delta Q\,.
\end{eqnarray}
Thus $T$ is directly linked  to the heat exchange, which is in turn related to the matter content.\\ 

It is worth to notice that here the variation of the heat, $\delta Q$, is considered with respect to the change in the volume, and not with respect to the ``thermal time'', as considered for the entropy production in Eq.~\eqref{SM:entropy production con Ricci target}. This allows us to identify
\begin{eqnarray}\label{SM:T=Q}
    \Big\langle\int d^4x\sqrt{-g}\,\,T\Big \rangle\propto-\frac{1}{\tilde{T}}\frac{\delta Q_{\text{matter}}}{\delta s}\,,
\end{eqnarray}
with $N$ lapse function and $\tilde{T}$ temperature.\\

Once we have interpreted the trace of the stress-energy tensor as the heat exchange induced by the matter content, we can consider that
\begin{eqnarray}\label{SM:R=Q+S}
    \Big \langle\int d^4x\sqrt{-g} R\, \Big\rangle\propto-\frac{1}{\tilde{T}}\frac{\delta Q_{\rm gravity}}{\delta s}+\frac{\delta S_{\rm gravity}}{\delta s}\,.
\end{eqnarray}
This identification allows us to regain the classic Clausius relation $\tilde{T}\delta S\geq\delta Q$. Moreover it is consistent with what found through the Smarr formula \cite{Wald,Smarr}, as it will be recovered in the next section.\\

Furthermore, the identification in Eq.~\eqref{SM:R=Q+S} seems to point out to the fact that $\tau=\Delta t$ in the integral over $d^4x$ is equal to ${\tilde{T}}^{-1}$. This has been already introduced in other approaches to thermodynamics --- see e.g. \cite{GH} --- and can be interpreted thanks to the intuitive argument presented in the previous section.\\

Therefore, considering the identification in Eq.~\eqref{SM:R=Q+S}, we see that the Ricci scalar contains both the gravitational entropy of the system and the heat exchange due to the gravitational processes --- as it can also be understood from the manipulation of the Komar mass in Eq.~\eqref{SM:Komar} that leads to the Smarr formula.

\section{The positivity relation}
\label{posit}
\noindent 
From the relation stating the positivity of the entropy we can extract an inequality between the Ricci scalar and the stress-energy tensor.\\

First, we notice that the positivity of the entropy induces an arrow of time with respect to the expansion (or better the contraction) of the Universe. In fact from Eq.~\eqref{SM:entropia e volume} we observe that
\begin{equation}
    \delta S\propto- \Big\langle\int d^4x \frac{1}{V_{\rm Pl}}\delta\sqrt{-g}\, \Big\rangle=- \Big \langle\frac{1}{V_{\rm Pl}}\delta V_{\rm manifold} \Big \rangle\,.
\end{equation}
We immediately infer that  
\begin{equation}\label{SM:freccia del tempo}
    \delta S\geq0 \rightarrow\langle\delta  V_{\rm manifold}\rangle\leq 0\,.
\end{equation}
From Eq.~\eqref{SM:freccia del tempo} we then recognize that a non-reversible transformation induces  a contraction (with respect to the thermal time) of the space-time volume.\\

It is important to notice that this change in volume does not regard any change of the integration extremes, but solely the change of the $\sqrt{-g}$ itself.\\

We may verify that Eq.~\eqref{SM:freccia del tempo} provides the right sign for the thermodynamical relation, once  Eq.~\eqref{SM:scalar Gauss}, which implies that $R^{(3)}+K^2-K_{ij}K^{ij}:=\mathcal{H}$, is plugged into Eq.~\eqref{SM:entropy production con Ricci target} ---  having neglected for simplicity the cosmological constant and using (for simplicity) a spacetime with shift function $\beta^i=0$ 
--- namely 
\begin{eqnarray}
    \frac{\delta S}{\delta s}\propto \Big \langle\int d^4x \sqrt{-g}\, (-2R_{\mu\nu}u^\mu u^\nu+\mathcal{H})\Big \rangle\geq0 \,.
\end{eqnarray}
If we suppose that a timelike Killing vector $k^\mu$ exists, require again for the sake of simplicity that $\beta=0$, and perform the saddle-point expansion, we obtain --- through a similar reasoning as the one explained in the previous section --- that
\begin{eqnarray}
    \int d^3x \sqrt{h}(-2R_{\mu\nu}u^\nu k^\mu)\geq -\int d^3x \sqrt{h}N(\mathcal{H})\,.
\end{eqnarray}

Considering a manifold with two boundaries, one at infinity, $\mathcal{S}$, and the other one on a Killing horizon $\mathrm{H}$, and choosing the normal vector suitably, we can observe  --- see e.g.~\cite{Wald} --- that the relation holds 
\begin{eqnarray}
    \int dV[-2R_{\mu\nu}u^\nu k^\mu]=\int_\mathcal{S} \alpha-\int_\mathrm{H} \alpha\,, 
\end{eqnarray}
with $\alpha_{ab}=\epsilon_{abcd}\nabla^ck^d$, as defined in \cite{Wald}.\\

The boundary term on $\mathcal{S}$ can be then interpreted as $-M$, total mass \cite{Wald}, while the boundary term on $\mathrm{H}$ can be demonstrated \cite{Wald} to be equal to 
\begin{eqnarray}
    -\int_\mathrm{H} \alpha=2\mathrm{k}A\,,
\end{eqnarray}
with $\mathrm{k}$ surface gravity and $A$ area of the horizon.\\

Therefore, combing all the previous relations, we obtain
\begin{eqnarray}\label{SM:Clausius gravitazionale}
    \frac{2}{8\pi}\mathrm{k}A=2TS\geq M-\int dVN \mathcal{H}:=Q_{gravity}\,,
\end{eqnarray}
where we interpreted $M-\int dVN\mathcal{H}$ as $U-W=Q$, with $W$ work --- with the sign convention for the work being the same one used in the Smarr formula --- and $Q$ heat. It then follows that the sign of the Clausius relation, $TdS\geq \delta Q$, is recovered.\\

During a non-reversible transformation, if the generator of the horizon is not the timelike Killing vector (as in stead it happens for a Kerr black hole, for example), the previous formula will give rise to additional terms that are related to the work, such as $\Omega_H J$, as demonstrated in \cite{Wald}.\\

We may now consider two thermal time values $s$ and $s_0$, where for $s_0$ the equality in Eq.~\eqref{SM:Clausius gravitazionale} holds. From Eq.~\eqref{SM:Clausius gravitazionale} we then obtain 
\begin{eqnarray}\label{SM:vero Clausius grav}
&&2T(s)S(s)=2T(s)S(s)-2T(s_0)S(s_0)+Q(s_0)\geq Q(s)\nonumber\\&&\longrightarrow\Delta^s_{s_0}(\frac{1}{4\pi}\mathrm{k}A)\geq \Delta^s_{s_0} Q \,.
\end{eqnarray}
The last inequality in Eq.~\eqref{SM:vero Clausius grav} provides us with a more general property than Eq.~\eqref{SM:freccia del tempo}, since it takes into account also the variation of the extremes of the integral, instead of a simple variation of $\sqrt{-g}$.\\

The form of the inequalities in Eq.~\eqref{SM:freccia del tempo} and Eq.~\eqref{SM:vero Clausius grav} may open new cosmological scenarios that go beyond Friedmann cosmology. Indeed, the Friedmann equations work solely for reversible transformations, being a consequence of the Einstein equations.\\

Finally, Eq.~\eqref{SM:freccia del tempo} provides us with an insightful definition of the arrow of time that may also represents a solution to the frozen formalism problem. Indeed, the frozen formalism can only be a consequence of the Einstein equations, thus of reversible transformations. From Eq.~\eqref{SM:entropy production con Ricci target} we finally recover, neglecting the cosmological constant, that
\begin{equation}
    \frac{\delta S}{\delta s}\propto\int Dx \, [R+kT]\, \geq 0\,.
\end{equation}
Since the latter relation holds for a generic volume, we can extrapolate the inequality
\begin{equation}
 \frac{\delta S}{\delta s}\geq0 \qquad \longrightarrow \qquad R\geq -kT\,, 
\end{equation}
which provides a lower bound for the Ricci scalar.

\section{An intuitive argument in favour of the Holographic principle}
\noindent
Let us consider again Eq.~\eqref{SM:R=Q+S}. Since the thermal time has physical dimension of the square of a length, the variation of the entropy with respect to the thermal time can be recast, in natural units $c=\hbar=G=k_B=1$, as
\begin{eqnarray}
    {\frac{\delta {S}_{\rm gravity}}{\delta s}}\sim \frac{1}{V_{\rm Pl}}\int d^4x \sqrt{-g}\, R\,.
\end{eqnarray}
The Einstein equations arise from the equivalence between the derivatives in the thermal time of the heat and the entropy. Thus the Planck volume cancels, and only the integral is left, having the dimensions of an area.\\

An intuitive argument about the origin of the area-law is the following --- for the sake of simplicity, we again assume the case $\beta^i=0$.\\

We start from the standard conservation law for the entropy
\begin{eqnarray}\label{SM:continuita_classica}
    \partial_t s=-\nabla\cdot J_s+\sigma\,,
\end{eqnarray}
with $s$ entropy density, $J_s$ entropy current and $\sigma$ entropy production. Adapting it in our case, we can write
\begin{eqnarray}
    \partial_s s=-\nabla_\mu J^\mu+\sigma\,.
\end{eqnarray}
Assuming now a spacetime with a timelike Killing vector, we obtain (writing the divergence explicitly)
\begin{eqnarray}
     &&\partial_s s=-\frac{1}{\sqrt{-g}}\partial_0(\sqrt{-g}J^0)-\frac{1}{\sqrt{-g}}(\partial_i  \sqrt{-g} J^i)+\sigma=\nonumber\\
     &&=-\frac{1}{\sqrt{-g}}(\partial_i  \sqrt{-g} J^i)+\sigma\,,
\end{eqnarray}
where in the last equality we used the fact that $\partial_0(\sqrt{-g}J^0)=0$,  the quantity being independent on time. Thus we find
\begin{eqnarray}
    \partial_s s=-\frac{1}{\sqrt{-g}}(\partial_i  \sqrt{-g} J^i)+\sigma\,,
\end{eqnarray}
which is the covariant version of the expression in Eq.~\eqref{SM:continuita_classica}.\\

Elaborating on this parallelism, we integrate in space the term $\nabla\cdot J$ 
and obtain the Newton's law for cooling, stating that
\begin{eqnarray}
    \int \,dV\, \nabla \cdot J_q=\int dA\, h\,\Delta T=h\,\Delta T\, A\,, 
\end{eqnarray}
with $A$ area of the boundary surface between the system and the environment, $h$ the heat transfer coefficient and $\Delta T$ the difference between the temperatures of the system and of the environment.\\

Recalling now that in absence of matter flow --- a reasonable assumption for a static black hole, in vacuum --- $J_s=-\frac{J_q}{T}$ (with $T$ temperature of the system \footnote{$T$ represents the local temperature. When it is realistic to consider the system to be at equilibrium, it has a constant temperature (given by vacuum fluctuations) in the whole space surrounding a black hole.}) we obtain, integrating the classical expression, that
\begin{eqnarray}
    \int dV \, \partial_t s=\int dV\, \frac{J_q}{T}+\int dV \, \sigma=\frac{A\, h\, \Delta T}{T}+\int dV \, \sigma\,.
\end{eqnarray}
Going back to the covariant expression, we find that
\begin{eqnarray}
    &&\frac{\delta S_{\rm gravity}}{\delta s}-\frac{\delta Q}{T\delta s}=\int d^4x \sqrt{-g}\, \partial_s s \, \,  -\frac{\delta Q}{T\delta s}=\nonumber\\
    &&=\int d^4x \sqrt{-g}\nabla_i J^i_s+\int d^4x \sqrt{-g}\sigma -\frac{\delta Q}{T\delta s}\, \propto\nonumber\\
    &&\propto\, \frac{1}{4\pi T}\mathrm{k}A-\frac{M}{T}+\frac{1}{T}\int dV NH\geq 0,\nonumber\\
\end{eqnarray}
where the proportionality relation arises from Eq.~\eqref{SM:Clausius gravitazionale}, having also used $\Delta t \propto \frac{1}{T}$.\\

We finally realize that: i) the area term arises directly from the general-relativistic formulation of the Newton's law of cooling; ii) the surface gravity $\mathrm{k}$, consequently, represents the difference between the temperature inside the black hole and  the temperature in the outer space; iii) the quantity $\int d^3x \sqrt{h}N H$ can be interpreted as the entropy production term of the out-of-equilibrium vacuum transformation --- namely, $h=\frac{1}{4\pi}$ within units $c=G=\hbar=k_B=1$.\\

This last interpretation is consistent with the fact that $H=0$ for reversible transformations in vacuum (i.e., for general relativity in the vacuum) and allows us to interpret the term $\int d^3x \sqrt{h}N H$ as out-of-equilibrium work (or dissipated gravitational heat).\\

For $\beta^i\neq0$, we can regard the term $2R_{\mu\nu}u^\mu \beta^\nu$ as a ``torsional'' dissipated heat, and then consider it as part of the entropy production, being it vanishing in the vacuum.

\bibliographystyle{apsrev4-1}
\bibliography{Reference}

@article{Barcelo_Liberati_Visser_2001,
  author        = {Barcel{\'o}, Carlos and Liberati, Stefano and Visser, Matt},
  title         = {Einstein gravity as an emergent phenomenon?},
  journal       = {International Journal of Modern Physics D},
  volume        = {10},
  pages         = {799--806},
  year          = {2001},
  eprint        = {gr-qc/0106002},
  archivePrefix = {arXiv},
  doi           = {10.1142/S0218271801001621}
}

@article{Chirco_Liberati_2010,
  author        = {Chirco, Goffredo and Liberati, Stefano},
  title         = {Non-equilibrium thermodynamics of spacetime: The role of gravitational dissipation},
  journal       = {Physical Review D},
  volume        = {81},
  pages         = {024016},
  year          = {2010},
  eprint        = {0909.4194},
  archivePrefix = {arXiv},
  primaryClass  = {gr-qc},
  doi           = {10.1103/PhysRevD.81.024016}
}

@article{Eling_2006,
  author        = {Eling, Christopher and Guedens, Raf and Jacobson, Ted},
  title         = {Non-equilibrium thermodynamics of spacetime},
  journal       = {Physical Review Letters},
  volume        = {96},
  pages         = {121301},
  year          = {2006},
  eprint        = {gr-qc/0602001},
  archivePrefix = {arXiv},
  doi           = {10.1103/PhysRevLett.96.121301}
}

@article{Jacobson_1995,
  author = {Jacobson, Ted},
  title = {Thermodynamics of Spacetime: The Einstein Equation of State},
  journal = {Phys. Rev. Lett.},
  volume = {75},
  pages = {1260--1263},
  year = {1995},
  doi = {10.1103/PhysRevLett.75.1260}
}

@article{Parisi_Wu_1981,
  author = {Parisi, Giorgio and Wu, Yong-Shi},
  title = {Perturbation Theory Without Gauge Fixing},
  journal = {Sci. Sin.},
  volume = {24},
  pages = {483},
  year = {1981}
}

@article{Onsager_Machlup_1953,
  author = {Onsager, Lars and Machlup, S.},
  title = {Fluctuations and Irreversible Processes},
  journal = {Phys. Rev.},
  volume = {91},
  pages = {1505--1512},
  year = {1953},
  doi = {10.1103/PhysRev.91.1505}
}

@article{Seifert_2012,
  author = {Seifert, Udo},
  title = {Stochastic Thermodynamics, Fluctuation Theorems and Molecular Machines},
  journal = {Rep. Prog. Phys.},
  volume = {75},
  pages = {126001},
  year = {2012},
  doi = {10.1088/0034-4885/75/12/126001}
}

@book{Gardiner,
  author = {Gardiner, C. W.},
  title = {Handbook of Stochastic Methods},
  publisher = {Springer},
  year = {2009}
}

@book{Risken,
  author = {Risken, H.},
  title = {The Fokker-Planck Equation},
  publisher = {Springer},
  year = {1996}
}

@article{MSR,
  author = {Martin, P. C. and Siggia, E. D. and Rose, H. A.},
  title = {Statistical Dynamics of Classical Systems},
  journal = {Phys. Rev. A},
  volume = {8},
  pages = {423--437},
  year = {1973},
  doi = {10.1103/PhysRevA.8.423}
}

@article{Lovelock,
    author ={D.Lovelock} ,
    title = {The Einstein Tensor and Its Generalizations},
    journal ={Journal of Mathematical Physics.} ,
    publisher={The American Institute of Physic},
    DOI={10.1063/1.1665613},
    volume={12},
   ISSN={1079-7114},
   number={3},
    year = {1971},
   month=march
}

@article{lulli2022stochasticquantizationgeneralrelativity,
      title={Stochastic Quantization of General Relativity \`a la Ricci-Flow}, 
      author={Matteo Lulli and Antonino Marciano and Xiaowen Shan},
      year={2025},
      journal = "Fortschritte der Physik",
      eprint={2112.01490},
      archivePrefix={arXiv},
      primaryClass={gr-qc},
      url={https://onlinelibrary.wiley.com/doi/10.1002/prop.70041}, 
}

@article{Sancho,
    author = {J. M. Sancho and M. San Miguel et al.},
    title = {Analytical and numerical studies of multiplicative noise},
    journal = {Physical Review A},
     volume={26},
     number={3},
    year = {1982},
    month=september
}

@article{Fox,
    author = {Ronald F.Fox},
    title = {Functional-calculus approach to stochastic differential equations},
    journal = {Physical Review A},
     volume={33},
     number={1},
    year = {1986},
    month=january
}

@book{Wald,
    author ={R. M. Wald} ,
    title = {General Relativity},
    publisher = {University of Chicago Press},
    year = {1984}
}

@article{York,
    author = {J.W York},
    title = {Boundary Terms in the Action Principles
of General Relativity},
    journal = {Foundations of Physics},
    volume={16},
    number={3},
    year = {1986}
}

@article{GH,
    author ={G. Gibbons, S. Hawking} ,
    title = {Action integrals and partition functions in quantum gravity },
    journal = {Physical Review D},
    volume={15},
    number={10},
    year = {1977}
}

@article{Smarr,
    author = {L. Smarr},
    title = {Mass Formula for Kerr Black Holes},
    journal = {Physical Review Letters},
    volume={129},
    number={4},
    year = {1973}
}

@article{DeDominicis,
  author = {De Dominicis, C.},
  title = {Techniques de Renormalisation de la Theorie des Champs et Dynamique des Phenomenes Critiques},
  journal = {J. Phys. Colloques},
  volume = {37},
  pages = {C1-247},
  year = {1976}
}

@article{Gurgolon,
    author = {E. Gourgoulhon},
    title = {3+1 Formalism and Bases of Numerical Relativity},
    journal = {Lecture Notes in Physics (Springer, Berlin)},
    volume={846},
    year = {2012}
}

@article{Novikov,
  author = {Novikov, E. A.},
  title = {Functionals and the Random-Force Method in Turbulence Theory},
  journal = {Sov. Phys. JETP},
  volume = {20},
  pages = {1290},
  year = {1965}
}

@article{Furutsu,
  author = {Furutsu, K.},
  title = {On the Statistical Theory of Electromagnetic Waves in a Fluctuating Medium},
  journal = {J. Res. Natl. Bur. Stand. D},
  volume = {67},
  pages = {303},
  year = {1963}
}

@article{Hamilton_1982,
  author = {Hamilton, Richard S.},
  title = {Three-Manifolds with Positive Ricci Curvature},
  journal = {J. Diff. Geom.},
  volume = {17},
  pages = {255--306},
  year = {1982}
}

@article{Perelman_2002,
  author = {Perelman, Grigori},
  title = {The Entropy Formula for the Ricci Flow and Its Geometric Applications},
  journal = {arXiv:math/0211159},
  year = {2002}
}

@article{Padmanabhan_2010,
  author = {Padmanabhan, T.},
  title = {Thermodynamical Aspects of Gravity: New Insights},
  journal = {Rep. Prog. Phys.},
  volume = {73},
  pages = {046901},
  year = {2010},
  doi = {10.1088/0034-4885/73/4/046901}
}

@article{Verlinde_2011,
  author = {Verlinde, Erik},
  title = {On the Origin of Gravity and the Laws of Newton},
  journal = {JHEP},
  volume = {1104},
  pages = {029},
  year = {2011},
  doi = {10.1007/JHEP04(2011)029}
}

@article{Rovelli_thermal_time,
  author = {Rovelli, Carlo},
  title = {Statistical Mechanics of Gravity and the Thermodynamical Origin of Time},
  journal = {Class. Quant. Grav.},
  volume = {10},
  pages = {1549--1566},
  year = {1993},
  doi = {10.1088/0264-9381/10/8/016}
}

@article{Crooks,
  title = {Path-ensemble averages in systems driven far from equilibrium},
  author = {Crooks, Gavin E.},
  journal = {Phys. Rev. E},
  volume = {61},
  issue = {3},
  pages = {2361--2366},
  numpages = {0},
  year = {2000},
  month = {Mar},
  publisher = {American Physical Society},
  doi = {10.1103/PhysRevE.61.2361},
  url = {https://link.aps.org/doi/10.1103/PhysRevE.61.2361}
}

@Article{Entropy,
AUTHOR = {Cates, Michael E. and Fodor, Étienne and Markovich, Tomer and Nardini, Cesare and Tjhung, Elsen},
TITLE = {Stochastic Hydrodynamics of Complex Fluids: Discretisation and Entropy Production},
JOURNAL = {Entropy},
VOLUME = {24},
YEAR = {2022},
NUMBER = {2},
ARTICLE-NUMBER = {254},
URL = {https://www.mdpi.com/1099-4300/24/2/254},
PubMedID = {35205548},
ISSN = {1099-4300},
DOI = {10.3390/e24020254}
}

\end{document}